\documentclass[twocolumn,showpacs,preprintnumbers,amsmath,amssymb]{revtex4}

\usepackage{graphicx}
\usepackage{dcolumn}
\usepackage{bm}

\begin{document}
\newcommand{\gdhi}{\ooalign{\hfil/\hfil\crcr$\partial$}}

\def\Sp{\mathop{\mathrm{Sp}}\nolimits}
\def\sgn{\mathop{\mathrm{sgn}}\nolimits}
\def\erfc{\mathop{\mathrm{erfc}}\nolimits}
\def\tr{\mathop{\mathrm{tr}}\nolimits}
\def\as{\mathop{\mathrm{as}}\nolimits}
\def\val{\mathop{\mathrm{val}}\nolimits}

\title{Quantization of the chiral soliton in medium}
% Force line breaks with \\

\author{S.Nagai}
\author{N.Sawado}
\email{sawado@ph.noda.tus.ac.jp}
\author{N.Shiiki}
\email{norikoshiiki@mail.goo.ne.jp}
\affiliation{
Department of Physics, Faculty of Science and Technology, 
Tokyo University of Science, Noda, Chiba 278-8510, Japan 
}
\date{\today}

\begin{abstract}
Chiral solitons coupled with quarks in medium are studied 
based on the Wigner-Seitz approximation. 
The chiral quark soliton model is used to obtain the classical
soliton solutions. To investigate nucleon and $\Delta$ in matter, 
the semi-classical quantization is performed by the cranking method. 
The saturation for nucleon matter and $\Delta$ matter are observed.   
\end{abstract}


\maketitle

\section{\label{sec:level1}Introduction\protect\\ } 
The study of dense nuclear matter with the internal nucleon structure 
is old but still a challenging subject. 
Especially, the approach of the topological soliton 
model seems interesting, because it is believed as a 
low energy effective model in the large $N_c$-limit of QCD.
It was first applied for nuclear matter system in 80's by using 
the skyrmion centered cubic (CC) crystal by Klebanov~\cite{klebanov85}. 
This configuration was studied further by W\"ust, Brown and Jackson 
to estimate the baryon density and discuss the phase transition 
between nuclear matter and quark matter~\cite{wust87}.  
Goldhabor and Manton found a new configuration, body centered cubic (BCC)   
of half-skyrmions in a higher density regime~\cite{manton87}. 
The face centered cubic (FCC) and BCC lattice were also 
studied by Castillejo {\it et al.}~\cite{castillejo89} 
and the phase transitions between those configurations were 
investigated by Kugler and Shtrikman~\cite{kugler89}. 
Recently, the idea of using crystallized skyrmions to study 
nuclear matter was revived by Park, Min, Rho and Vento 
with the introduction of the Atiyah-Manton multi-soliton ansatz 
in a unit cell~\cite{park02}. 

Incorporating quark degrees of freedom into each soliton 
makes the prediction more realistic. 
Achtzehnter, Scheid and Wilets investigated the Friedberg-Lee 
soliton bag model with a simple cubic lattice~\cite{achtzehnter85}. 
Due to the periodicity of the background potential, the solution of 
the Dirac equation has the form of the Bloch waves, 
$\psi_{\bm k}({\bm r})=e^{i\bm{k}\cdot\bm{r}}\phi_{\bm k}({\bm r})$ 
where $\phi_{\bm k}$ satisfies the same periodic boundary condition 
as the background potential. They performed the calculation for 
only one direction of the crystal momenta ${\bm k}=k{\bm e_z}$ and 
assumed the spherically symmetric energy surface. 
The Bloch condition is, however, anisotropic for the nonzero ${\bm k}$ 
and the results should be highly dependent on the approximation. 
The analysis of the crystal soliton 
model with quarks based on the Wigner-Seitz approximation 
has been already done. In this ansatz, a single soliton 
is placed on the center of a spherical unit cell. 
Then the lowest energy level (``bottom'' of the band) for the valence quarks becomes 
s-state. The appropriate boundary conditions at the cell 
boundary should be imposed on the quark wave functions 
as well as the chiral fields. This simple treatment sheds 
light on the nucleon structure in nuclear medium. Soliton 
matter within this approximation have been extensively 
studied by using various nucleon models such as the 
the chiral quark-meson type model
~\cite{banerjee85,glendenning86,hahn87,weber98}, 
Friedberg-Lee soliton bag model~\cite{reinhardt85,weber98,birse88,barnea00}, 
the Skyrme model~\cite{Kutschera84}.
The non-zero dispersion of the lowest band \cite{weber98} 
and the quark-meson coupling \cite{barnea00} were also examined within this approximation. 

The chiral quark soliton model (CQSM) can be interpreted as the soliton bag model 
including not only valence quarks but also the vacuum sea quark polarization 
effects explicitly~\cite{diakonov88,reinhardt88,meissner89}. The model provides 
correct observables of a nucleon such as mass, electromagnetic
value, spin carried by quarks, parton distributions
and octet, decuplet $SU(3)$ baryon spectra~\cite{report}. Remarkably  
this model predicted the exotic quark bound state, pentaquark 
$\Theta^+$~\cite{diakonov97} which was successfully observed 
in experiments~\cite{nakano03}.

Amore and De Pace studied soliton matter in the CQSM using the Wigner-Seitz 
approximation and observed the nuclear saturation~\cite{amore00}. 
They examined the soliton solutions with three different boundary  
conditions imposed on the quark wave function. However the obtained 
saturation density was higher than the experimental value and they 
concluded that such discrepancy is originated in the approximate    
treatment of the sea quark contribution~\cite{adjali92}. 
Thus we treat the vacuum polarization exactly in the manner 
originally proposed by Kahana and Ripka~\cite{kahana84} and semi-classically 
quantize the chiral soliton by the cranking method to see those effects 
on the matter solution. 
At present, soliton matter has been studied only at the classical energy level.  
In order to study the property of nucleon or $\Delta$ in medium, 
the spin and isospin of each of the soliton must be quantized. 
We hence perform the rotational collective quantization by the cranking 
formula and observe the saturation of nuclear and $\Delta$ matter. 
As shown in Sec. V, we obtained very shallow saturation. 

Unfortunately, the study of the nuclear matter within the soliton 
model often fails to fit the experimental values, even in the saturation energy.
This may be caused by the fact that the topological soliton picture is based 
on the approximation of large $N_c$-limit of QCD and therefore works well 
only in the very low energy scale. 
Thus our model improves slightly the situation in the sense that we take 
into account the quantum correction of $O(1/N_c)$ to the classical soliton 
mass of $O(N_c)$. However, it should be noted that as our model contains 
the valence quark explicitly, the physical meaning of such $N_c$ counting 
is obscure. 
Of course, the prescription is still insufficient, and the obtained results 
will still room for improvement. 

\section{\label{sec:level2}The chiral quark soliton model\protect\\ }
The CQSM was originally derived from the instanton 
liquid model of the QCD vacuum and incorporates the non-perturbative 
feature of the low-energy QCD, spontaneous chiral symmetry breaking.  
The vacuum functional is defined by~\cite{diakonov88}
\begin{eqnarray}
	{\cal Z} = \int {\cal D}\pi{\cal D}\psi{\cal D}\psi^{\dagger}\exp \left[ 
	i \int d^{4}x \, \bar{\psi} \left(i\!\!\not\!\partial
	- MU^{\gamma_{5}}\right) \psi \right]	 \label{vacuum_functional}
\end{eqnarray} 
where the SU(2) matrix
\begin{eqnarray}
	U^{\gamma_{5}}= \frac{1+\gamma_{5}}{2} U + \frac{1-\gamma_{5}}{2} U^{\dagger} 
\end{eqnarray}
with
\begin{eqnarray}
	U=\exp \left( i \bm{\tau} \cdot \bm{\phi}/f_{\pi} \right)
	=\frac{1}{f_\pi}(\sigma+i\bm{\tau}\cdot\bm{\pi})
\end{eqnarray}
describes chiral fields, $\psi$ is quark fields and $M$ is the dynamical 
quark mass. We choose the constituent quark mass $M=420$ MeV which reproduces 
the experimental observables of the free nucleon correctly~\cite{report}. 
$f_{\pi} $ is the pion decay constant and experimentally 
$f_{\pi} \sim 93 {\rm MeV}$. 
Since our concern is the tree-level pions and one-loop quarks according 
to the Hartree mean field approach, the kinetic term of the pion fields which 
gives a contribution to higher loops can be neglected. 
Due to the interaction between the valence quarks and the Dirac sea, 
soliton solutions appear as bound states of quarks in the background of self-consistent 
mean chiral field. $N_{c}$ valence quarks fill the each bound state to form a baryon. 
The baryon number is thus identified with the number of bound states filled by 
the valence quarks \cite{kahana84}. 
The $B=1$ soliton solution has been studied in detail at classical and 
quantum level in \cite{diakonov88,reinhardt88,meissner89,
report,wakamatsu91}.  

The vacuum functional in Eq.(\ref{vacuum_functional}) can be integrated 
over the quark fields to obtain the effective action 
\begin{eqnarray}
	S_{{\rm eff}}[U]&=&-iN_{c}{\rm lndet}\left(i
	\!\!\not\!\partial - MU^{\gamma_{5}}\right)\label{effective_action1}\\
	&=&-\frac{i}{2}N_{c}{\rm Spln}D^{\dagger}D
	\label{effective_action2}
\end{eqnarray}
where $i D=i\gdhi - MU^{\gamma_{5}}$~($iD$ is called the Dirac operator).
This determinant is ultraviolet divergent and must be 
regularized. Using the proper-time regularization scheme, we can write 
\begin{eqnarray}
	&&S^{{\rm reg}}_{{\rm eff}}[U]
	=\frac{i}{2}N_{c}
	\int^{\infty}_{1/\Lambda^2}\frac{d\tau}{\tau}{\rm Sp}\left(
	{\rm e}^{-D^{\dagger}D\tau}-{\rm e}^{-D_{0}^{\dagger}D_{0}\tau}\right) 
	\nonumber \\
	&&=\frac{iN_{c}T}{2}\int^{\infty}_{-\infty}\frac{d\omega}{2\pi}
	\int^{\infty}_{1/\Lambda^2}\frac{d\tau}{\tau}{\rm Sp}\left[{\rm e}
	^{-\tau (H^2+\omega^2)}-{\rm e}^{-\tau (H_{0}^2+\omega^2)}\right] 
	\label{regularised_action} \nonumber \\
\end{eqnarray}
where $T$ is the Euclidean time separation, $\Lambda$ is a cut-off 
parameter evaluated by the condition that the derivative expansion of 
Eq.(\ref{effective_action1}) reproduces the pion kinetic term with the 
correct coefficient ${\it i.e.}$
\begin{eqnarray}
	f_{\pi}^2=\frac{N_{c}m^2}{4\pi^2}\int^{\infty}_{1/\Lambda^2} 
	\frac{d\tau}{\tau}{\rm e}^{-\tau M^2}
	\,\, , \label{cutoff_parameter}
\end{eqnarray}
and $H$ is the Dirac one-quark Hamiltonian defined by
\begin{eqnarray}
	H=\frac{\bm{\alpha}\!\cdot\!\nabla}{\it{i}}+\beta M 
	U^{\gamma_{5}}\,\,. \label{hamiltonian}
\end{eqnarray}
$D_{0}\equiv D(U=1)$ and $H_{0}\equiv H(U=1)$ correspond to 
the vacuum sectors. 
At $T \rightarrow \infty$, we have ${\rm e}^{iS_{{\rm eff}}}
\sim  {\rm e}^{-iE_{\rm sea}T}$. Integrating over $\omega$ 
in Eq.(\ref{regularised_action}) and constructing a complete set of 
eigenstates of $H$ with 
\begin{eqnarray}
	H|\nu\rangle = \epsilon_{\nu}|\nu\rangle\,\, ,\,\,\,
	H_{0}|\nu\rangle^{(0)} = \epsilon^{(0)}_{\nu}|\nu\rangle^{(0)}
	\,\, , \label{eigen_equation}
\end{eqnarray}
one obtains the sea quark energy \cite{meissner89}
\begin{eqnarray}
	E_{\rm sea}[U]=\frac{N_{c}}{4\sqrt{\pi}}\int^{\infty}_{1/\Lambda^2}
	\frac{d\tau}{\tau^{3/2}}\left(\sum_{\nu}{\rm e}^{-\tau\epsilon_{\nu}^2}
	-\sum_{\nu}{\rm e}^{-\tau\epsilon^{(0)2}_{\nu}}\right).
	\nonumber \\
	\label{energy_sea}
\end{eqnarray}
In the Hartree picture, the baryon states are the quarks occupying all 
negative Dirac sea and valence levels. Hence, if we define the total soliton energy 
$E_{\rm static}$, the valence quark energy $E_{\rm val}[U]$ should be added;
 \begin{eqnarray}
	E_{\rm static}[U]=N_{c}E_{\rm val}[U]+E_{\rm sea}[U]\,\,. 
	\label{energy_static}
\end{eqnarray}
To obtain the $B=1$ soliton solution, we impose the hedgehog ansatz on the chiral field  
\begin{eqnarray}
U(\bm{r})=\exp(i F(r) \hat{\bm{r}}\cdot \bm{\tau})=\cos F(r)+i\hat{\bm{r}}\cdot \bm{\tau}\sin F(r)
\label{chiral_fields_hedgehog}
\end{eqnarray}
with the boundary conditions 
\begin{eqnarray}
F(0)=-\pi,~~F(\infty)=0\,.
\label{boundary_condition}
\end{eqnarray}
The one-quark hamiltonian (\ref{hamiltonian}) reads
\begin{eqnarray}
H(U^{\gamma_5})=-i\alpha\cdot\nabla + \beta M(\cos F(r)+i\gamma_5\hat{\bm{r}}\cdot 
\bm{\tau}\sin F(r))\,. \nonumber \\
	\label{hamiltonian_hedgehog}
\end{eqnarray}
This hamiltonian does not commute with the total angular momentum $\bm{j}$ nor the 
isospin $\bm{\tau}/2$ but commute with the grand spin operator 
$\bm{K}=\bm{j}+\bm{\tau}/2$ with $[H,\bm{K}]=0$. 
$H$ also commutes with the ${\cal P}=\gamma_0$ which turns to be a parity operator. 
As a result, the one-quark eigenstates are labeled by the $K=0,1,2,\cdots$ and the 
parity ${\cal P}=\pm$. The three valence quarks occupy the  
lowest states $K^{\cal P}=0^{+}$ and are responsible for the baryon number $(=1)$ 
(non topological charge). 
 
Field equations for the chiral fields can be obtained by demanding  
that the total energy in Eq.(\ref{energy_static}) be stationary 
with respect to variation of the profile function $F(r)$,
\begin{eqnarray*}
	\frac{\delta}{\delta F(r)}E_{\rm static}=0 \,\, ,
\end{eqnarray*}
which produces  
\begin{eqnarray}
	S(r)\sin F(r)=P(r)\cos F(r),  
	\label{field_equation}
\end{eqnarray}
where 
\begin{eqnarray}
S(r)&=&N_{c}\sum_\nu\bigl(n_\nu\theta(\epsilon_\nu)+{\rm sign}(\epsilon_\nu)
{\mathcal N}(\epsilon_\nu)\bigr) \nonumber \\
&\times&\langle \nu |\gamma^{0}\delta(|x|-r)|\nu\rangle\,, 
\\	
P(r)&=&N_{c}\sum_\nu\bigl(n_\nu\theta(\epsilon_\nu)+{\rm sign}(\epsilon_\nu)
{\mathcal N}(\epsilon_\nu)\bigr) \nonumber \\
&\times&\langle \nu |i \gamma^{0}\gamma^{5}\hat{\bm{r}}
\cdot\bm{\tau}\delta(|x|-r)|\nu\rangle \, .
\end{eqnarray}
with 
\begin{eqnarray*}
{\cal N}(\epsilon_{\nu})=-\frac{1}{\sqrt{4\pi}}\Gamma\left(\frac{1}{2},
\left(\frac{\epsilon_\nu}{\Lambda}\right)^2\right)
\end{eqnarray*}
and $n_\nu$ is the valence quark occupation number.

The procedure to obtain the self-consistent solution of Eq.(\ref{field_equation}) 
is that $1)$ solve the eigenequation in (\ref{eigen_equation}) under an assumed 
initial profile function $F_{0}(r)$, $2)$ use the resultant eigenfunctions and 
eigenvalues to calculate $S(r)$ and $P(r)$, $3)$ solve 
Eq.(\ref{field_equation}) to obtain a new profile function, $4)$ repeat $1)-3)$ 
until the self-consistency is attained.

\begin{figure}
\includegraphics[height=7cm, width=9cm]{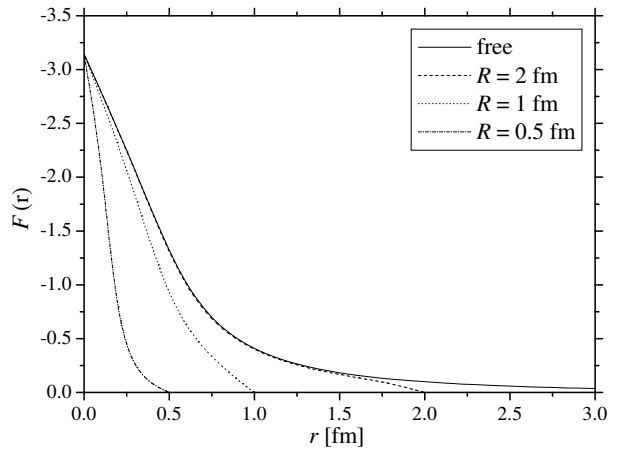}
\caption{\label{fig:Fig1} Profile functions for $R=0.5,1,2$ fm and the 
free soliton. }
\end{figure}

\section{\label{sec:level4}The numerical basis\protect\\ }
In this section we present the numerical method of eigen problem of the hamiltonian~(\ref{hamiltonian_hedgehog}).
The hamiltonian with hedgehog ansatz commutes with the parity and the grandspin operator 
given by  
\begin{eqnarray*}
	\bm{K}=\bm{j}+\bm{\tau}/2=\bm{l}+\bm{\sigma}/2+\bm{\tau}/2,
\end{eqnarray*}
where $\bm{j},\bm{l}$ are respectively total angular momentum and orbital angular momentum. 
Accordingly, the angular basis can be written as  
\begin{eqnarray}
|(lj)KM\rangle= \sum_{j_3\tau_3}C^{KM}_{jj_3\frac{1}{2}\tau_3}
\Bigl(\sum_{m\sigma_3}C^{jj_3}_{lm\frac{1}{2}\sigma_3}
|lm \rangle |\frac{1}{2}\sigma_3 \rangle \Bigr) |\frac{1}{2} \tau_3 \rangle\,.\nonumber\\
\end{eqnarray}
For $B=1$ solution, following states are possible 
\begin{eqnarray}
&&|0\rangle =|(K~K+\frac{1}{2})KM \rangle\,,  \nonumber   \\
&&|1\rangle =|(K~K-\frac{1}{2})KM \rangle\,,  \nonumber   \\
&&|2\rangle =|(K+1 K+\frac{1}{2})KM\rangle\,, \nonumber  \\
&&|3\rangle =|(K-1 K-\frac{1}{2})KM\rangle\,. \nonumber
\end{eqnarray}
With this angular basis, the normalized eigenstates of the free hamiltonian 
in a spherical box with radius $R$ can be constructed as follows:
\begin{eqnarray}
&&u^{(a)}_{KM}=
N_k\left( 
\begin{array}{c}
i\omega^{+}_{\epsilon_k}j_{K}(kr)|0\rangle \\
\omega^{-}_{\epsilon_k}j_{K+1}(kr)|2\rangle
\end{array}
\right), \nonumber \\
&&u^{(b)}_{KM}=
N_k\left( 
\begin{array}{c}
i\omega^{+}_{\epsilon_k}j_{K}(kr)|1\rangle \\
-\omega^{-}_{\epsilon_k}j_{K-1}(kr)|3\rangle
\end{array}
\right), \nonumber \\
&&v^{(a)}_{KM}=
N_k\left( 
\begin{array}{c}
i\omega^{+}_{\epsilon_k}j_{K+1}(kr)|2\rangle \\
-\omega^{-}_{\epsilon_k} j_{K}(kr)|0\rangle
\end{array}
\right), \nonumber \\
&&v^{(b)}_{KM}=
N_k\left( 
\begin{array}{c}
i\omega^{+}_{\epsilon_k}j_{K-1}(kr)|3\rangle \\
\omega^{-}_{\epsilon_k} j_{K}(kr)|1\rangle
\end{array}
\right), 
\label{kahana_ripka}
\end{eqnarray}
with
\begin{eqnarray}
	N_k=\biggl[\frac{1}{2}R^3
	\Bigl(j_{K+1}(kR)\Bigr)^2\biggr]^{-1/2}
\end{eqnarray}
and $\omega^{+}_{\epsilon_k>0},\omega^{-}_{\epsilon_k<0}={\rm sgn}(\epsilon_k), 
\omega^{-}_{\epsilon_k>0},\omega^{+}_{\epsilon_k<0}=k/(\epsilon_k+M)$.
The  $u$ and $v$ correspond to the {\it ``natural''} 
and {\it ``unnatural''} components of the basis  
which stand for parity $(-1)^{K}$ and $(-1)^{K+1}$ respectively. 
The momenta are discretized by the boundary conditions $j_K(k_i R)=0$. 
The orthogonality of the basis is then satisfied by  
\begin{eqnarray}
&&\int^R_0 dr r^2 j_K(k_i r)j_K(k_j r) \nonumber \\
&&=\int^R_0 dr r^2 j_{K\pm 1}(k_i r)j_{K\pm 1}(k_j r)  \nonumber \\
&&=\delta_{ij}\frac{R^3}{2}  [j_{K\pm 1}(k_i R)]^2 \, .
\label{orthogonality}
\end{eqnarray}

\begin{table}
\caption{\label{tab:comp_clamass}The classical mass 
for the original Kahana-Ripka basis and modified version (in MeV), 
with $M=400$ MeV, $R=6$ fm. The error becomes of order $\sim 10^{-3}$.}
\begin{ruledtabular}
\begin{tabular}{cccc}
      & Valence & Vacuum & Total \\ \hline 
free  & 191 & 637  & 1209 \\
modified & 192 & 633 & 1210 \\
\end{tabular}
\end{ruledtabular}
\end{table}

Let us examine the boundary conditions for the chiral and Dirac fields 
to construct the nuclear matter solution in the Wigner-Seitz approximation.
When the background chiral fields are periodic with lattice vector $\bm{a}$, 
the quark fields would be replaced by Bloch wave functions as  
$\psi(\bm{r}+\bm{a})=e^{i\bm{k}\cdot \bm{a}}\psi(\bm{r})$. 
In the Wigner-Seitz approximation, however, the soliton is put on the 
center of the spherical unit cell with the radius $R$ ($a=2R$) 
and the dispersion $\bm{k}$ is assumed to be zero.
For the profile function $F(r)$, the periodicity and the unit topological 
charge inside the cell require the boundary conditions 
\begin{eqnarray}
\left.
\begin{array}{c} 
~\sigma'(0)=\sigma'(R)=0   \\
\pi(0)=\pi(R)=0
\end{array}\right\} \Rightarrow F(0)=-\pi, F(R)=0\,.
\end{eqnarray}
For the Dirac eigenstates, modification in the basis is needed. 
For odd number of $K$, the boundary condition is same as the free case with 
\begin{eqnarray}
j_K(k_i R)=0\,.
\end{eqnarray}
For even $K$, the following conditions must be satisfied 
\begin{eqnarray}
j_{K+1}(k^{(a)}_i R)=0,~~{\rm for}~~ u^{(a)}_{KM},v^{(a)}_{KM}\,,\nonumber \\
j_{K-1}(k^{(b)}_i R)=0,~~{\rm for}~~ u^{(b)}_{KM},v^{(b)}_{KM}\,.
\label{mboundary}
\end{eqnarray}
Obviously the conditions (\ref{mboundary}) partially break  
the orthogonality of the basis (\ref{orthogonality}) for the finite value of $R$.  
However we can solve the eigenvalue problem properly (see Table \ref{tab:comp_clamass}). 
Although the motivation is different, the similar treatment has been already introduced 
in Ref.\cite{weigel92}.

\begin{figure}
\includegraphics[height=7cm, width=9cm]{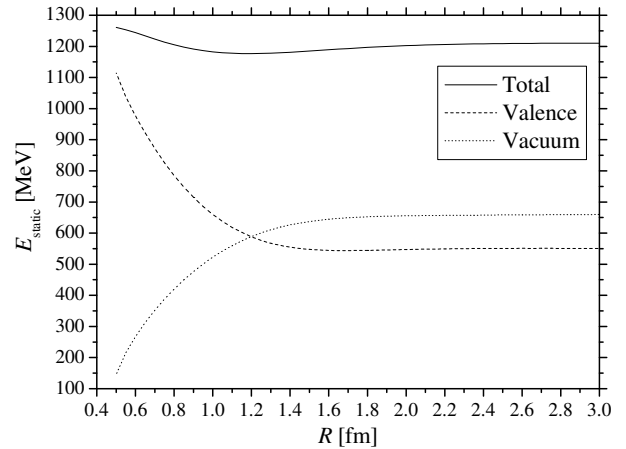}
\caption{\label{fig:Fig2} Classical soliton energy and its valence and 
vacuum contributions~(\ref{energy_static}).}
\end{figure}

Fig.~1 shows the self-consistent profile functions for free ($R\to \infty$) and 
various values of the cell radius $R$. In Fig.~2, we present the results of the 
classical energy of the soliton and its valence and vacuum contributions
as functions of $R$. We find the shallow minimum of the classical energy
at $R\sim 1.2$ fm. 

\section{\label{sec:level5}Spurious center of mass correction\protect\\ }
\begin{figure}
\includegraphics[height=7cm, width=9cm]{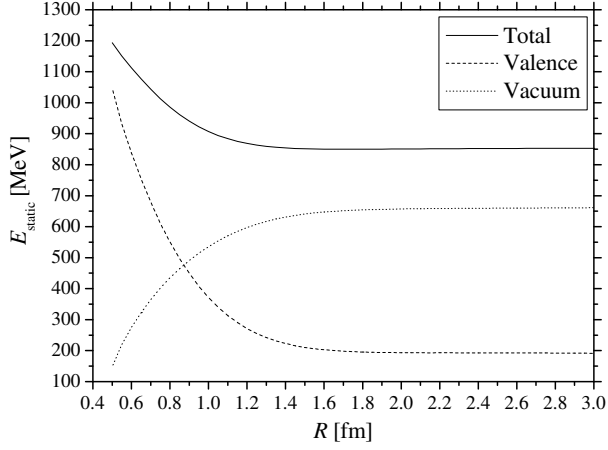}
\caption{\label{fig:Fig3} Classical soliton energy after removing the spurious center 
of mass motion~(\ref{cmcorrect}). }
\end{figure}

\begin{figure}
\includegraphics[height=7cm, width=8cm]{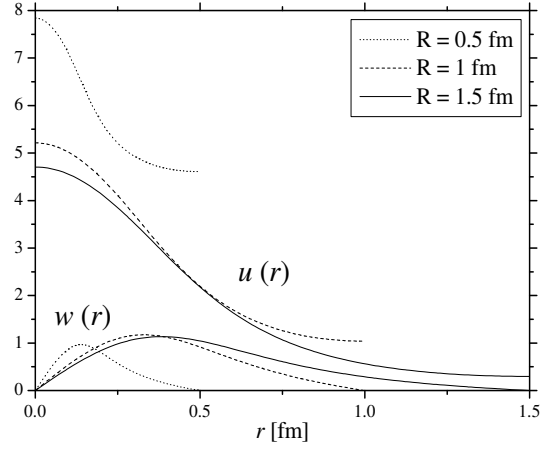}
\caption{\label{fig:Fig4} The ``upper'' $u(r)$ and the ``lower'' $w(r)$ 
component of valence quark wave functions for various cell
radius $R$ with the boundary condition $w(R)=0$.
Non vanishing values of upper component at the cell boundary $u(R)$ 
come from the zero-mode elements in the basis. }
\end{figure}

The minimum found in Fig.~\ref{fig:Fig2} is not regarded as a true saturation 
point because it contains the zero-point energy contributions. The quark 
contribution to the mean-field expectation value of the square of the 
total momentum $\bm{P}^2$ appears at the classical level although it should 
be zero because the soliton is rest at the cell center in the present 
approximation. Therefore, the corresponding kinetic energy should be 
subtracted from the total energy. The effects of the spurious center 
of mass motion is roughly estimated by the method of 
Pobylitsa {\it et al.}~\cite{pobylitsa92}. Considering the translational 
degrees of freedom and performing their quantization, one obtains 
the correction at a rest frame as  
\begin{eqnarray}
E_{\rm  static}\to\tilde{E}_{\rm static}= E_{\rm static}-\frac{\langle \bm{P}^2\rangle}{2 E_{\rm static}}\,.
\label{cmcorrect}
\end{eqnarray}
The correction  is easily evaluated by using the  
numerical basis given in Eq.~(\ref{kahana_ripka}) as  
$\bm{P}^2 u^{(a)}_{KM}(k_i r)=k_i^2 u^{(a)}_{KM}(k_i r)$.
As is shown in Fig.~\ref{fig:Fig3}, the minimum disappears after removing the 
zero-point energy contributions~(\ref{cmcorrect}). 
This is explained by the observation that the contribution of the center of 
mass motion becomes small with increasing density (see Fig.~\ref{fig:Fig4} and the caption).  

\section{\label{sec:level6}Collective quantization\protect\\ }
The solitons that we have obtained in the previous section are 
classical objects and therefore must be 
quantized to assign definite spin and isospin to them.  
For the solitons in the free space, quantization can be 
performed semiclassically for their rotational zero modes. 
For the hedgehog soliton, because of its topological structure, 
a rotation in isospin space is followed by a simultaneous spatial 
rotation. Let us introduce the dynamically rotated chiral fields 
\begin{eqnarray}
	\tilde{U}(\bm{x},t)=A(t)U(\bm{x})A(t)^{\dagger},~~	
	A(t)\in {\rm SU}(2)_I\,.
	\label{cranking}
\end{eqnarray}
In a crystal configuration, the solitons are fixed on the spatial lattice 
point and their isospin orientation is chosen so as to minimize 
the energy of the system. If one rotates each soliton in the crystal, 
it changes the isospin orientation and increases the energy. 
Thus there is only one isospin collective coordinate corresponding 
to the overall orientation of the crystal in isospace, called global isospin, 
in the soliton crystal~\cite{klebanov85,baskerville96}.  

The Wigner-Seitz treatment with spherical cell approximation 
may cure the situation. Because in this approximation the information of 
the crystalline structure, hence, the isospin structure is completely lost 
at least in the low-density, the rotational zero-mode would be recovered. 
Thus, we apply the zero-mode quantization method to the WS-cell 
to estimate the nucleon and the delta mass splitting in the matter. 

By transforming the rotating frame of reference, the Dirac operator 
with Eq.~(\ref{cranking}) can be written as 
\begin{eqnarray}
	\tilde{iD}= A(t){\gamma}^0 [i {\partial}_t -
	H(U^{{\gamma}_{5}}) + \Omega]A(t)^{\dagger} 
\end{eqnarray}
where 
\begin{eqnarray}
	\Omega=i{A^{\dagger}} \dot{A}=\frac{1}{2} \Omega^a \tau_a\,.
\end{eqnarray}
$\Omega$ is the angular velocity operators for an isorotation. 
Assuming that the rotation of the soliton is adiabatic, 
we shall expand the effective action $S_{\rm eff}$ around the classical 
solution $U(\bm{x})$ with respect to the angular momentum velocity 
$\Omega$ up to second order~\cite{biedenharn85} 
\begin{eqnarray}
	&&S_{\rm eff}(\tilde{U}) = S_{\rm eff}(U) \nonumber \\
	&&-iN_c\Sp\left[
	\log \bigl(i {\partial}_t-H 
	+ \Omega\bigr)\right]
	-\Sp\left[\log (i {\partial}_t-H)\right].\nonumber 
\end{eqnarray}
With the proper-time regularization, we have 
\begin{eqnarray}
	S^{\rm reg}_{\rm eff}(\tilde{U})= S^{\rm reg}_{\rm eff}(U) \nonumber
	+\frac{1}{2} \sum_{ab}\int dt \bigl[
	  I_{{\rm sea},ab} \Omega^{a}(t) \Omega^{b}(t) \bigr]
\end{eqnarray}
where $I_{{\rm sea},ab}$ is the vacuum sea contribution to the moments of inertia
defined by 
\begin{eqnarray}
	I_{{\rm sea},ab} = \frac{1}{8}N_c \sum_{\nu,\mu}f(\epsilon_\mu,\epsilon_\nu,\Lambda) 
	{\langle \nu|{\tau}_a | \mu \rangle} 
	{\langle \mu|{\tau}_b |\nu \rangle}\,, 
	\label{inertiao}
\end{eqnarray}
with the cutoff function $f(\epsilon_\mu,\epsilon_\nu,\Lambda)$ 
\begin{eqnarray}
	&&f(\epsilon_\mu,\epsilon_\nu,\Lambda)
	=-\frac{2\Lambda}{\sqrt{\pi}}
	\frac{e^{-\epsilon_\mu^2/\Lambda^2}-e^{-\epsilon_\nu^2/\Lambda^2}}{\epsilon_\mu^2-\epsilon_\nu^2} 
	\nonumber \\
	&&+\frac{{\rm sgn}(\epsilon_\mu){\rm erfc}(|\epsilon_\mu|/ \Lambda)
	-{\rm sgn}(\epsilon_\nu){\rm erfc}(|\epsilon_\nu|/ \Lambda)}{\epsilon_\mu-\epsilon_\nu}\,. \nonumber 
\end{eqnarray}
Similarly, for the valence quark contribution we have  
\begin{eqnarray}
	I_{{\rm val},ab} = \frac{1}{2}N_c \sum_{\mu \neq \val} 
	\frac{{\langle \val|{\tau}_a | \mu \rangle} {\langle \mu|{\tau}_b |\val \rangle}}
	{E_\mu - E_{\rm val}}\,. 	
	\label{inertiav}
\end{eqnarray}
The total moments of inertia are then given by the sum of the vacuum 
and valence, $I_{ab} = I_{{\rm val},ab} + I_{{\rm sea},ab}$.
The hedgehog ansatz of the chiral fields ensure the relation for 
the moment of inertia
\begin{eqnarray}
I_{11}=I_{22}=I_{33}.
\end{eqnarray}
The quantization condition for the collective coordinate, $A(t)$, 
define a body-fixed isospin operator $\bm{K}$ as 
\begin{eqnarray}
	&&I_{ab} {\Omega}^{b} \rightarrow -
	\tr \bigg( A \frac{ {\tau}_a }{2} \frac{\partial}{\partial A}  \biggr) 
	\equiv -k_a\,, 	
	\label{qcondition}
\end{eqnarray}
These are related to the usual coordinate-fixed isospin 
operator $i_a$ by transformation,
\begin{equation}
	i_{a}=- \frac{1}{2}{\rm Tr}[\tau_aA(t)\tau^b A(t)^{\dagger}]k_b.
	\label{orthotrans}
\end{equation}       
To estimate the quantum energy corrections, let us introduce 
the basis functions of the spin and isospin operators which
were inspired from the cranking method for 
nuclei~\cite{bohr},
\begin{eqnarray*}
	&&\langle A|{i{i}_{3}{k}_{3}}\rangle
	=\sqrt{\frac{2i+1}{8\pi^2}}(-1)^{i+i_3}
	D^{i}_{-i_{3}k_{3}}(A) 
\end{eqnarray*}  
where $D$ is the Wigner rotation matrix. Finally, we find the 
quantized energies of the soliton as 
\begin{eqnarray}
	&&E=E_{\rm static}+\frac{i(i+1)}{2I_{33}}
	\label{qenergy}
\end{eqnarray}
where $i(i+1)$ is eigenvalues of the Casimir operator 
$\bm{i}^{2}$.
\begin{figure}
\includegraphics[height=7cm, width=9cm]{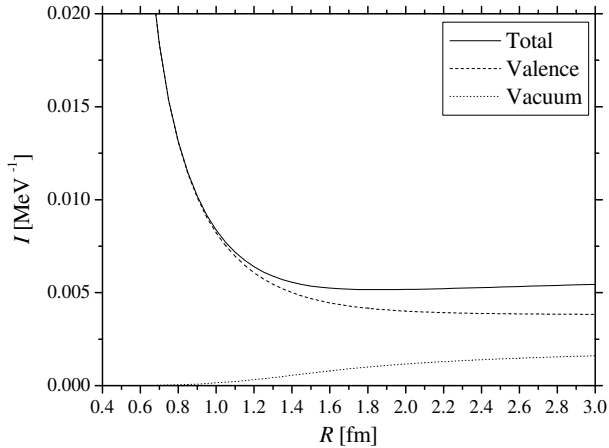}
\caption{\label{fig:Fig5} Moment of inertia : the vacuum
 (\ref{inertiao}) and the valence (\ref{inertiav}) 
contribution and their sum. }
\end{figure}
The moment of inertia for the vacuum (\ref{inertiao}) 
and valence (\ref{inertiav}) and their sum are given in Fig.~\ref{fig:Fig5}. 
In Fig.~\ref{fig:Fig6}, we present the energy of 
nucleon ($i=\frac{1}{2}$) and $\Delta$ ($i=\frac{3}{2}$).  

In this cranking procedure, the zero-point energy of the 
rotational motion $ \langle\bm {i^2}\rangle / 2I_{33}
$ must be removed from Eq.~(\ref{qenergy})~\cite{cohen86,pobylitsa92}. 
Finally, we obtain the mass of nucleon and delta 
\begin{eqnarray}
E_N=\tilde{E}_{\rm static}-\frac{3}{4I_{33}}\,, \label{qmassn} \\
E_\Delta=\tilde{E}_{\rm static}+\frac{3}{4I_{33}}\,. \label{qmassd}
\end{eqnarray}
Fig.~\ref{fig:Fig7} shows the energy of nucleon and delta after 
subtracting the spurious zero-point energy. 
The minimum for nucleon is observed at $R\sim 1.8$ fm which corresponds 
to the density $\rho_N \sim 0.04~{\rm fm}^{-3}$. This value is much lower 
than the experimental value.
The binding energy is $E_{B}\sim 18$ MeV which is not 
far from the experimental observation. 
For $\Delta$, we also find the shallow minima at $R\sim 1.22$ fm
which corresponds to $\rho_\Delta \sim 0.13~{\rm fm}^{-3}$. 
The $\Delta$ saturation is attained at the density $\rho_\Delta/\rho_N\sim 3.2$ 
which is close to the prediction of density $\rho_\Delta/\rho_N\sim 2-3$
in the framework of the quantum hadrodynamics~\cite{waldhauser87,li97}.
The advantage of our approach is that the model does not require any 
tuning parameter for the $\Delta$ spectra in the hadrodynamics calculations.

\begin{figure}
\includegraphics[height=7cm, width=9cm]{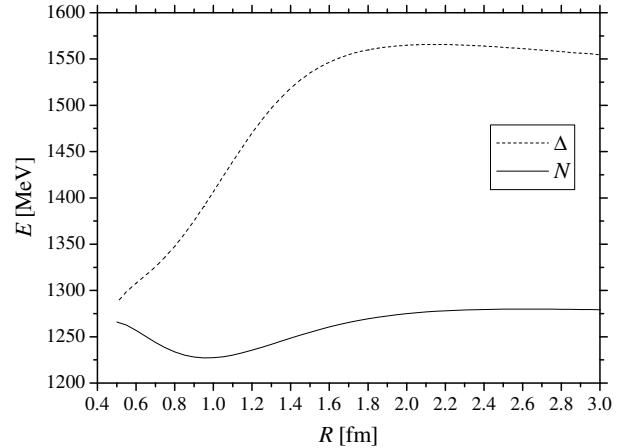}
\caption{\label{fig:Fig6} Quantized soliton energies of nucleon $N$
and delta resonance $\Delta(1232)$~(\ref{qenergy}).}
\end{figure}

\section{\label{sec:level7}Summary\protect\\ }
We have studied soliton solutions in nuclear medium by using the Wigner-Seitz 
approximation. The chiral quark soliton model was used to obtain the classical 
soliton solution. In this letter we especially focused on the properties of nucleon 
and $\Delta$ in matter. We quantized the soliton semiclassically. The adiabatic 
rotation for the (iso-)rotational zero mode was performed  
and the nuclear saturation points were obtained for nucleon and $\Delta$ matter. 

\begin{figure}
\includegraphics[height=7cm, width=9cm]{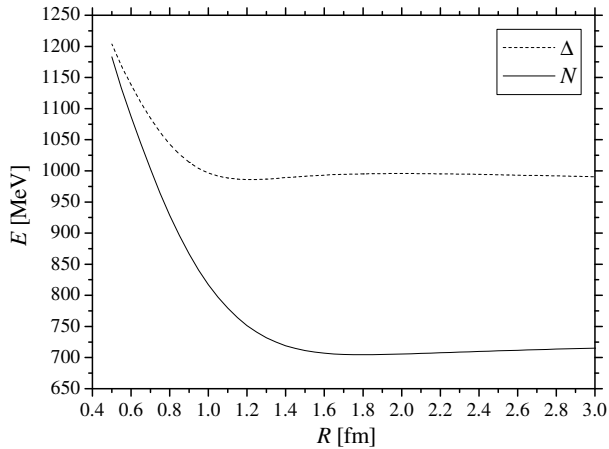}
\caption{\label{fig:Fig7} Masses of $N,\Delta$, 
after spurious energy subtractions~(\ref{qmassn})-(\ref{qmassd}).}
\end{figure}

\begin{figure}
\includegraphics[height=7cm, width=9cm]{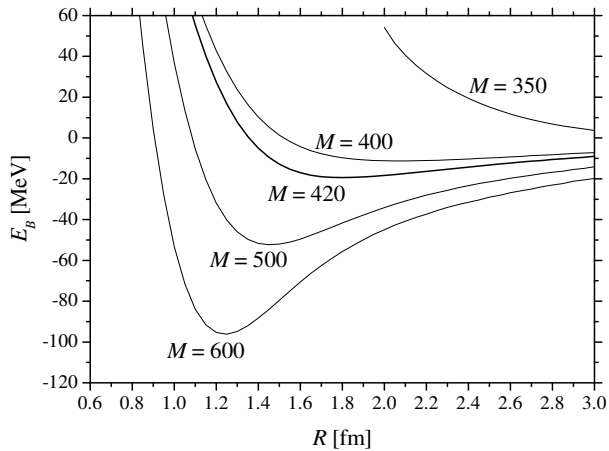}
\caption{\label{fig:Fig8} Binding energy of nucleon for  
the various constituent quark mass $M$ (in MeV).}
\end{figure}

\begin{figure} 
\includegraphics[height=7cm, width=9cm]{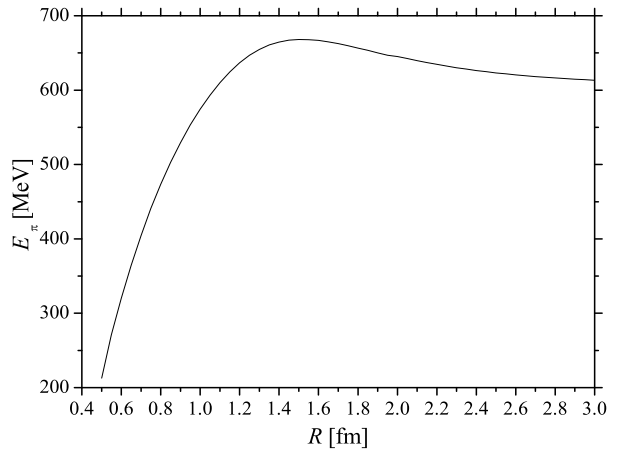} 
\caption{\label{fig:Fig9} Kinetic energy of the pion~(\ref{kineticpi}).} 
\end{figure}
 
Here we did not consider the following effects 
which should be investigated to develop our understanding of the dense nuclear matter:
\begin{itemize}
\item band structure of the quarks 
\item $R$ dependence of the constituent quark mass $M$ and the cutoff parameter 
for the vacuum $\Lambda$
\item inclusion of the heavier mesons ($\rho,\omega,\cdots$) to the soliton solutions
\item improvement of the correction by the zero point energy and Casimir effects
\item quark-meson couplings and the Fermi motion of the baryons
\item crystalline order in high density phase
\item $SU(3)$ extension. 
\end{itemize}
As is expected, our model provides much lower value of the saturation 
density than the experiment. In this analysis, the Wigner-Seitz cell is 
approximated by a sphere and thus high density matter is attained 
by shrinking the cell volume with the spherical shape of each soliton 
unchanged. However, in reality, the neighborhood solitons start to overlap 
and the structure will be deformed from uniform nuclear matter at high density. 
In this phase, the hedgehog ansatz should not be appropriate any more. 

We observed the increase in the zero mode of the center of mass motion 
of the soliton for higher density, which means that 
the soliton tends to rest in the WS approximation. 
In this case, we should employ the exact WS cell which reflects 
the background crystal symmetry instead of sphere, to get higher 
saturation density. 
The inclusion of band effects may also improve our results. 
In Ref.~\cite{barnea00}, the authors imposed the Bloch-like 
boundary conditions on the s-wave valence quark wave function 
and estimated the soliton energy self-consistently. 
They found that the effects of the admixtures of higher states are 
small except for the scalar quark density. 
In fact, the band structure will appear at some critical density 
and the correction for the quantum energy may become more 
important at the dense medium because the radius of the soliton, that is, 
the moment of inertia, strongly depend on the position of the band~\cite{amore00}.

Generally speaking, the constituent quark mass $M$ is momentum- and 
density-dependent~\cite{diakonov88}. We chose the value $M=420$ MeV as 
it reproduces the free nucleon observable. 
In Fig.~\ref{fig:Fig8}, one can see that for larger value of $M$, 
the saturation point goes to inward and the binding becomes deeper. 
Varying the value of $M$ for each density may give a better result for 
the saturation point. 

An important feature of the nucleon in a matter is about its size. It is believed that 
the nucleon will swell in the medium. 
The authors of Ref.~\cite{christov93} observe such effect 
with reducing effective quark mass $M^{*}$ in the Nambu-Jona-Lasinio type quark-soliton model.
We confirmed within our model that 
as smaller the $M$, the size of the soliton increases. 
But in that case, the saturation becomes shallow (Fig.~\ref{fig:Fig8}). 
Recently, we investigated soliton solutions in the CQSM taking into 
account $\rho,\omega$ mesons which will improve the short distance physics. 
We are able to obtain deeper binding energy as decreasing the value of $M$. 
We will report it on forthcoming article. 

In Fig.~\ref{fig:Fig6}, one finds that the spectra of nucleon and  $\Delta$ 
are too small compared to the experimental values. Obviously it is due to the 
subtraction of the zero-point corrections. A little more sophisticated approach of 
the spurious motion is performed in Ref.~\cite{barnea00} and by applying 
this approach to our analysis, the results will be improved to a certain extent. 
Also, the meson coupling to the quark inside nucleon and $\Delta$ should be  
important to shift the minima at higher density.  

In Fig.~\ref{fig:Fig6} and Fig.~\ref{fig:Fig7}, one finds the nucleon-$\Delta$ 
mass difference gradually decreases as matter density increases and eventually  
it vanishes. The reduction in the mass difference has been observed 
previously in a similar chiral soliton model but employing somewhat 
different projection technique for quantum number~\cite{arriola89, christov90}. 
In the present formulation, the behavior is not fully understood 
because it should be explained by the dynamics of hadrons, that is, QCD. 
In the naive $SU(6)$ quark model, the mass difference is ascribed to 
the hyperfine splitting~\cite{glashow75}. The reduction may imply 
the increase of the distance between quarks. 
In fact, in Fig.~\ref{fig:Fig4}, one can see the concentration of 
the quarks at the cell boundary as the density increases.

Alternatively, if we understand the $\Delta$ as a composite object (resonance state) 
of the nucleon and pion, the mass difference can be interpreted as the energy of pions 
bound to the nucleon. Although it is absent in the present formulation, 
the pion kinetic energy inside the soliton can be estimated as
\begin{eqnarray} 
E_{\pi}&=&\frac{f_\pi^2}{4}\int d^3x {\rm tr}\partial_k U^\dagger \partial_k U \nonumber \\ 
&=& 2\pi f_{\pi}^2\int^R_0 r^2dr\biggl(F'(r)^2+\frac{\sin^2 F(r)}{r^2}\biggr). 
\label{kineticpi} 
\end{eqnarray}
In Ref.~\cite{amore00}, the authors introduced the $r$- and the cutoff parameter of the vacuum 
$\Lambda$-dependent form of the pion decay constant $f_\pi(r,\Lambda)$ 
and estimated its density dependence with the $\Lambda$ whose value is set 
for the free space value of $f_\pi$. 
The $f_\pi(r,\Lambda)$ determined in such a way is essentially valid only for the 
free space limit $R\to\infty$.
Therefore we shall simply take the value in free space $f_{\pi}=93$ MeV. 
Fig.~\ref{fig:Fig9} shows the result of the kinetic energy of pions 
as a function of $R$ and one can observe that the energy is 
reduced as the density increases. This reduction of the pion kinetic energy 
may contribute to the reduction of the mass difference. 

Our formulation is directly applicable to the $SU(3)$ octet-decuplet baryon 
spectra in nuclear matter~\cite{weigel92,blotz93}. 
After the above effects are properly incorporated and more
realistic estimation of the saturation points is achieved, 
it will be interesting to study the $SU(3)$.


\begin{thebibliography}{qq}
\bibitem{klebanov85}
Igor Klebanov, Nucl. Phys. B {\bf 262}, 133 (1985).
\bibitem{wust87}
E. W\"ust, B. E. Brown and A. D. Jackson, 
Nucl. Phys. A {\bf 468}, 450 (1987).
\bibitem{manton87}
Alfred S. Goldhaber and N. S. Manton, 
Phys. Lett. B {\bf 19}, 231 (1987).
\bibitem{castillejo89}
L. Castillejo, P. S. Jones, A. D. Jackson, 
J. J. M. Verbaarschot and A. Jackson, 
Nucl. Phys. A {\bf 501}, 450 (1987).
\bibitem{kugler89}
M. Kugler and S. Shtrikman, 
Phys. Rev. D {\bf 40}, 3421 (1989).
\bibitem{park02}
Byung-Yoon Park, Dong-Pil Min, Mannque Rho and 
Vincente Vento, Nucl. Phys. A {\bf 707}, 381 (2002).
\bibitem{achtzehnter85}
Joachim Achtzehnter, Werner Scheid and Lawrence Wilets, 
Phys. Rev. D {\bf 32}, 2414 (1985). 
\bibitem{banerjee85}
B. Banerjee, N. K. Glendenning and V. Soni, 
Phys. Lett. B {\bf 155}, 213 (1985).
\bibitem{glendenning86}
N. K. Glendenning and B. Banerjee, 
Phys. Rev. C {\bf 34}, 1072 (1986).
\bibitem{hahn87}
Detlev Hahn and Norman K. Glendenning, 
Phys. Rev. C {\bf 36}, 1181 (1987).
\bibitem{weber98}
Urban Weber and Judith A. McGovern, 
Phys. Rev. C {\bf 57}, 3376 (1998). 
\bibitem{reinhardt85}
H. Reinhardt, B. V. Dang, and H. Schulz, 
Phys. Lett. B {\bf 159}, 161 (1985).
\bibitem{birse88}
M. C. Birse, J. J. Rehr and L. Wilets, 
Phys. Rev. C {\bf 38}, 359 (1988).
\bibitem{barnea00}
Nir Barnea, Timothy S. Walhout, 
Nucl. Phys. A {\bf 677}, 367 (2000).
\bibitem{Kutschera84}
M. Kutschera, C. J. Pethick and D. G. Ravenhall, 
Phys. Rev. Lett. {\bf 53}, 1041 (1984).
\bibitem{diakonov88}
D. I. Diakonov, V. Yu. Petrov, and P. V. Pobylitsa, 
Nucl. Phys. B {\bf 306}, 809 (1988).
\bibitem{reinhardt88}
H. Reinhardt and R. W\"unsch , Phys. Lett. {\bf B 215,} 577 (1988).
\bibitem{meissner89}
Th. Meissner, F. Gr\"ummer, and K. Goeke, 
Phys. Lett. B {\bf 227}, 296 (1989).
\bibitem{report} For detailed reviews of the model see: \\
R.\ Alkofer, H.\ Reinhardt and H.\ Weigel, Phys.\ Rept.\ {\bf 265}, 139 (1996);\\
 Chr.\ V.\ Christov, A.\ Blotz, H.-C.Kim, P.\ Pobylitsa, T.\ Watabe, Th.\ Meissner, 
E.\ Ruiz Arriola, K.\ Goeke, Prog.\ Part.\ Nucl.\ Phys.\ {\bf 37}, 91 (1996).
\bibitem{diakonov97} D. Diakonov, V. Petrov and M. Polyakov, Z. Phys. A 
{\bf 359}, 305 (1997).
\bibitem{nakano03} T. Nakano {\it et al.}, Phys. Rev. Lett. {\bf 91}, 012002
 (2003).
\bibitem{amore00}
P. Amore and A. De Pace, 
Phys. Rev. C {\bf 61}, 055201 (2000).
\bibitem{adjali92}
I. Adjali, I. J. Aitchison, and J. A. Zuk, 
Nucl. Phys. A {\bf 537}, 457 (1992).
\bibitem{kahana84}
S. Kahana and G. Ripka, Nucl. Phys. A {\bf 429}, 462 (1984).
\bibitem{wakamatsu91}
M. Wakamatsu and H. Yoshiki, Nucl. Phys. A {\bf 524}, 561 (1991).
\bibitem{weigel92}
H. Weigel, R. Alkofer and H. Reinhardt, 
Nucl. Phys. B {\bf 387}, 638 (1992).
\bibitem{pobylitsa92}
P. V. Pobylitsa, E. Ruiz Arriola, Th. Meissner, F. Gr\"ummer, 
K. Goeke and W. Broniowski, J. Phys. G {\bf 18}, 1455 (1992).
\bibitem{baskerville96}
W.K.Baskerville, Phys. Lett. B {\bf 380}, 106 (1996).
\bibitem{biedenharn85}
L. C. Biedenharn, Y. Dothan and M. Tarlini, 
Phys. Rev. D {\bf 31}, 649 (1985).
\bibitem{cohen86}
Thomas D. Cohen and Wojciech Broniowski, 
Phys. Rev. D {\bf 34}, 3472 (1986).
\bibitem{bohr}
A. Bohr and B. Mottelson, {\it Nuclear structure, Vol.II}
(World Scientific Publishing Co. Pte. Ltd, Singapore, 1998).
\bibitem{waldhauser87}
B. M. Waldhauser, J. Theis, J. A. Maruhn, H. St\"ocker, and
W. Greiner, 
Phys. Rev. C {\bf 36}, 1019 (1987).
\bibitem{li97}
Zhuxia Li, Guangjun Mao, Yizhong Zhuo, and Walter Greiner, 
Phys. Rev. C {\bf 56}, 1570 (1997).
\bibitem{christov93}
Chr.V.Christov, K.Goeke, 
Nucl. Phys. A {\bf 564}, 551 (1993).
\bibitem{arriola89}
E.Ruiz Arriola, Chr.V.Christov and K. Goeke, 
Phys. Lett. B {\bf 225}, 22 (1989).
\bibitem{christov90}
Chr.V.Christov, M. Fiolhais, E.Ruiz Arriola and K. Goeke, 
Phys. Lett. B {\bf 243}, 333 (1990).
\bibitem{glashow75}
A. De Rujula, H. Georgi and S. L. Glashow, 
Phys. Rev. D {\bf 75},147 (1975).
\bibitem{blotz93} 
A.\ Blotz, D.\ Diakonov, K.\ Goeke, N.\ W.\ Park, V.\ Petrov
and P.\ V.\ Pobylitsa, Nucl.\ Phys.\ A {\bf 555}, 765 (1993).
\end{thebibliography}
\end{document}